\documentclass[11pt]{article}

\usepackage{color,helvet,times}
\usepackage{setspace} 
\usepackage{amsmath}
\usepackage{amssymb}
\usepackage{graphicx}
\usepackage{multirow}
\usepackage{hyperref}

\usepackage{amscd}
\usepackage{chemarrow}
\usepackage{verbatim}
\usepackage[normalem]{ulem}

\usepackage{geometry} 
\newcommand{\md}{d\kern-0.035cm\char39\kern-0.08cm}
\newcommand{\mL}{L\kern-0.15cm\char39}
\def\ml{l\kern-0.035cm\char39\kern-0.03cm}
\def\mt{t\kern-0.035cm\char39\kern-0.03cm}
\def\mL{L\kern-0.08cm\char39}

\newcommand{\eq}[1]{\ref{#1}}

\usepackage{calrsfs}

\renewcommand{\subsubsection}[1]{\paragraph{#1.}}

\begin{document}

\title{\bf Emerging Polynomial Growth Trends in COVID-19 Pandemic Data and Their Reconciliation with Compartment Based Models}
\author{Katar{\'\i}na  Bo{\md}ov\'a, Richard Koll\'ar\\ 
\it  Faculty of Mathematics, Physics and Informatics, Comenius University, \\Mlynsk\'a Dolina, 84248 Bratislava, Slovakia\\
(Dated: May 13, 2020)
}

\date{}
\pagestyle{myheadings}

\maketitle

\begin{abstract}
We study the reported data from the COVID-19 pandemic outbreak in January - May 2020 in 119 countries. 
We observe that the time series of active cases in individual countries (the difference of the total number of confirmed infections and the sum of the total number of reported deaths and recovered cases)  display a strong agreement with polynomial growth and at a later epidemic stage also with a combined polynomial growth with exponential decay. Our results are also formulated in terms of  compartment type mathematical models of epidemics. Within these models the universal scaling characterizing the observed regime in an advanced epidemic stage can be interpreted as an algebraic decay of the relative reproduction number $R_0$ as $T_M/t$, where $T_M$ is a constant and $t$ is the duration of the epidemic outbreak.  We show how our findings can be applied to improve predictions of the reported pandemic data and estimate some epidemic paramters.  Note that although the model shows a good agreement with the reported data we do not make any claims about the real size of the pandemics as the relation of the observed reported data to the total number of infected in the population is still unknown. 
\end{abstract}


\section{Introduction}
The coronavirus disease 2019 (COVID-19) pandemic caused by the severe acute respiratory syndrome coronavirus 2 (SARS-CoV-2) is accompanied by an unprecedented challenge for its mathematical modeling. Most of the difficulties stem in an extremely high level of uncertainty in available data:
\begin{itemize}
\item
Methodology for reporting of all types of data (number of confirmed positive cases, hospitalizations, recovered individuals, and even confirmed deaths) is not systematic and varies from one country to another \cite{COVIDMethodology}. 
\item
Different types of tests and test protocols used for COVID-19 detection have their own limitations both in sensitivity and specificity; testing procedures differ in methodology of a sample selection and in a testing sample size in different countries. In addition, different types of tests detect different phases of individual infections and their results differ based on the clinical stage of the infection \cite{BSImmun}.  
\item
A relation of the reported data to the real (unobserved) number of infected in a population is not understood and the estimates for the ratio of total cases in population to the number of observed cases vary even in the order of magnitude \cite{Manski2020, Fuhrmann2020, Ferguson, BBCNewsMerkel, SantaClaraCOVID, SantaClaraANTICOVID, VoCOVID, AustriaCOVID}.
\end{itemize}
The extent of uncertainty in data prohibits designing and validating mathematical models that would provide verifiable  accurate description of dynamics of the pandemic. That in turn has serious consequences on pandemic spread control and efficient epidemiological decision making. Particular difficulty for mathematical modeling is that traditional compartment type models, often referred to as ``SIR models" with Susceptible -- Infected -- Recovered compartments, that succeeded in an accurate description of previous epidemics, produce scenarios that do not match closely the observed data \cite{Ferguson, IHME} and fail to capture significant trends observed in data, see also \cite{Kucharski, Wu} for other early models. While it is still too early to call whether these models fail to capture the real dynamics of pandemics (as the observed data may not completely correspond to number of infected in the whole population) there is an urgent  need to understand the available data and its relation to the SIR  models.

\subsection{Note of precaution}
\label{s:NP}
We consider very important to emphasize the limited scope of our analysis as in the current literature some of these limitations are blurred and may eventually lead to misinterpretations. To overcome the difficulties with mathematical modeling and a high degree of uncertainty in data we propose a simple  descriptive model that captures the dynamics of the observed data rather than a detailed mechanistic model of the pandemic dynamics in the whole population. Therefore our results need to be interpreted with high precaution. We only present a systematic mathematical description of the observed reported data. We provide neither an explanation of the pandemics spreading, nor any claim about the real scope of the pandemics. However, if one conjectures that the observed reported data capture the extent of the pandemic size in the real population (e.g., if data systematically report a fixed percentage of the total infected population), then our results can be used to identify the nature of the emerging trends in pandemics spreading in individual countries. Furthermore, we only model the first wave of the pandemic and we do not make any predictions about the extent and timing of next waves, although we discuss how our method can be used to study effect of next epidemic waves. 
We are aware that we ignore multiple key factors that may influence the relation between the observed data and the real epidemic size (the level of detection of infected by testing, delays in test reporting, variation in clinical aspects of the infection, etc.). Note that in the text we use the term COVID-19 for all reported infections based on a positive PCR test and thus we do not distinguish between the presence of the virus in the respiratory tract of an infected and the disease it causes. 
Through the whole text we model the time series of the total number of active cases that is the difference of the reported total number of confirmed infections and the sum of the total number of reported deaths and recovered cases.

\subsection{Our work}
We have identified  a universal trend in the data reported by individual countries that helps to improve predictions for the final observed epidemic size and the related time scales. The universal trend is observed despite inhomogeneities and uncertainties in the data time series and various types and levels of mitigation policies applied. There is a transition in time series from an exponential growth (EG) to a polynomial growth (PG) and at a later stage to a combined polynomial growth with an exponential decay (PGED) in the number of active cases across almost all countries world-wide (with a sufficient size of current data). Our choice of the form of the trend in data is motivated by the theoretical results of \cite{Vazquez, Szabo, Maier,Brandenburg} and by the data analysis performed in \cite{Ziff,Li} that relate the observed transition to structural changes in the population contact networks. 

We analyze the reported data and estimate the parameters for individual countries. This allows us to categorize the countries into groups according to their advance through the first wave of the pandemics, and also to detect a possible divergence into a possible upcoming second wave in some countries. We rank a selected group of countries according to one of model parameters that captures their ability to identify, test, and isolate infected individuals from the rest of the population and thus prevent further spreading.   We also provide a reconciliation of the PGED regime with the SIR model that allows to build PGED directly into existing SIR models despite the fact that PGED is inconsistent with the SIR models (in their traditional form).   The PGED regime corresponds to an explicit algebraic decay of the relative reproduction number $R_0$  in time $t$  in the form $R_0 \approx T_M/t$ with a constant $T_M$. The trend is in general agreement with the current estimates of the evolution of $R_0$ in many individual countries \cite{Epiforecasts}; see also \cite{He} for a data supporting the observed dependence. 
A model based on PGED regime can also aid forecasting of the future dynamics of the observed data including an analysis of eventual next epidemic waves.

\section{Compartment Based Model in Epidemiology (SIR Model)}
The keystones of mathematical modeling of epidemic spread are the compartment based models also referred to as SIR models introduced by Kermack and McKendrick \cite{KMcK1, KMcK2,KMcK3}, see also \cite{MurrayI,Brauer} for an extensive  literature overview. We only review basic properties of the SIR model relevant for our further analysis. 

The basic SIR model describes the dynamics of susceptible ($S = S(t)$), infected ($I = I(t)$), and recovered ($R = R(t)$) populations at time $t$ by the coupled system of ordinary differential equations
\begin{eqnarray}
\frac{dS}{dt} &=& - \beta \frac{S}{N} \, I\, , \label{Seq}\\
\frac{dI}{dt} & = & \phantom{+} \beta \frac{S}{N}\, I - \gamma I \, , \label{Ieq}\\
\frac{dR}{dt} & = & \phantom{+} \gamma I\, . \label{Req}
\end{eqnarray}
Here $N$ is the total population, $\beta >0$ is the infection rate, and $\gamma >0$ is the removal rate of infections. The key assumptions behind this mechanistic model are 
\begin{itemize}
\item
the total population $N = S + I + R$ is constant,
\item
the population is well mixed, thus likelihood of a contact and transmission of  the infection from any infected to any susceptible is the same,
\item
the populations $S$, $I$ and $R$ are large enough to be well approximated by the real variable instead of integers. 
\end{itemize}
The basic model can be readily extended to account for an incubation period (SEIR model), for age structured and geographically structured populations, etc. Also, the assumption of the constant total population can be removed. However, the well mixed population assumption cannot be completely removed unless one considers SIR models on networks. In that case a good knowledge of the underlying contact network is necessary to calibrate the model to any type of data.

\subsection{Relative reproduction number}
The current extent of epidemic spread in the system is typically measured by the relative reproduction number $R_0 = R_0(t)$ that is related to the instantaneous rate of growth of the infected population. Equation \eq{Ieq} can be written as 
$dI / dt = \gamma (R_0 - 1) I$ for 
$$
R_0 = \frac{\beta}{\gamma} \, \frac{S}{N} \, .
$$
Thus  $R_0 = 1$ corresponds to the epidemic peak, the tipping point that characterizes the moment when the population of infected individuals is stationary, $dI / dt = 0$.%
\footnote{In epidemiology literature the basic reproduction number $R_0$ is typically defined as $R_0 = \beta/ \gamma$ and then the basic reproduction number that is constant during the epidemic outbreak is compared with the relative size of the susceptible population $s  = S/N$ to quantify the present speed of spreading.}
Note that $dI / dt >0$ for $R_0> 1$ and $dI/dt < 0$ for $R_0 < 1$ (for $I > 0$).
For practical reasons related to  the COVID-19 outbreak it is useful to rewrite the expression for $R_0$ as 
\begin{equation}
R_0 = \beta \, s\, T_{inf}\, ,
\label{R0def}
\end{equation}
Here $s = s(t) = S(t)/N$  is the proportion of the infected in the population at time $t$. The time scale $T_{inf}$ is associated with the removal of an average individual from the infected population, i.e., the typical length of the period during which the infected individual infects the susceptible population. It  is related  to $\gamma$ by $T_{\inf} = 1/\gamma$.  The infection rate $\beta$ can be expressed as $\beta = p/T_{sus}$, where $T_{sus}$ is the typical time scale associated with occurrence of interactions between susceptible and infected in the population where infection transfer may happen. The total number of transmission contacts between susceptible and infected population is given by $sI/T_{sus}$. Finally, $p$ is the probability of infection of a susceptible individual through a single random meeting with an infected. 

\subsection{Epidemic control}
In the SIR model  the number of active cases decreases due to a decrease of $R_0$ below 1. As $\beta$ and $T_{inf}$ are constant it is achieved through a reduction of the proportion of the susceptible population $s$ below the threshold $1/(\beta T_{inf})$. That in general requires a significant decrease of the susceptible population through  vaccination, gained natural immunity, infection  (herd immunity) or through a long term quarantine of a large fraction of the susceptible population.  However, the pandemic size  apparently has not reached such a high level yet. Thus $R_0$ needs to be controlled through a decrease in $\beta$ or $T_{inf}$. The parameter $\beta$ is an obvious candidate as strict public mitigation measures and social distancing decrease $p$ and increase $T_{sus}$. These measures are typically associated with a large economic cost and also parameters of this structural change on the level of the contact network are still unknown and their direct effect on $R_0$ cannot be accurately quantified.  A decrease of the parameter $T_{inf}$ does not require widespread mitigation measures. Although ability to decrease $T_{inf}$ solely based on the observation of disease symptoms   is limited from below by the length of the incubation period,  $T_{inf}$ can be significantly reduced by active contact tracing.

\section{Exponential Growth Regime (EG)}
\label{s:EG}
Until mitigation measures are applied we expect that an epidemic outbreak is governed by the SIR model \eq{Seq}--\eq{Req}. 
That implies an exponential growth (EG) of $I = I(t)$ during the early stages of epidemic. For completeness we present the asymptotic behavior of $(S,I,R)$ in \eq{Seq}--\eq{Req} for $t\rightarrow 0^+$.  
Let $R_{00} = R_0(0)$ and $(S(0), I(0),R(0)) = (S_0, I_0, 0)$. Then $S(t) \approx S_0$ for $t \ll 1$ and \eq{Ieq} reduces to 
$dI / dt \approx \gamma (R_{00} - 1) I$, i.e., 
\begin{equation*}
I(t) \approx I_0 \exp\Big(\gamma(R_{00}-1)t\Big)\, .
\label{IapproxExp}
\end{equation*}  
Consequently from \eq{Req} and $S+ I + R = N$ we obtain
$$
R(t) \approx  \frac{I_0}{R_{00}-1} \left[\exp \Big( \gamma(R_{00}-1)t \Big) -1 \right]\, , \ \ 
S(t)  \approx N - \frac{R_{00}}{R_{00}-1} I_0 \exp \Big(\gamma (R_{00}-1)t\Big) \, , 
$$
that is consistent with \eq{Seq}.  Note that EG was not observed in data in multiple countries (e.g., Slovakia, Lithuania) as these countries introduced mitigation measures at very early stages of the epidemic (zero or only a few confirmed cases of infection).

\section{Polynomial Growth Regime (PG)}
\label{s:PG}
\begin{figure}
\centering
\includegraphics[width=0.95\textwidth]{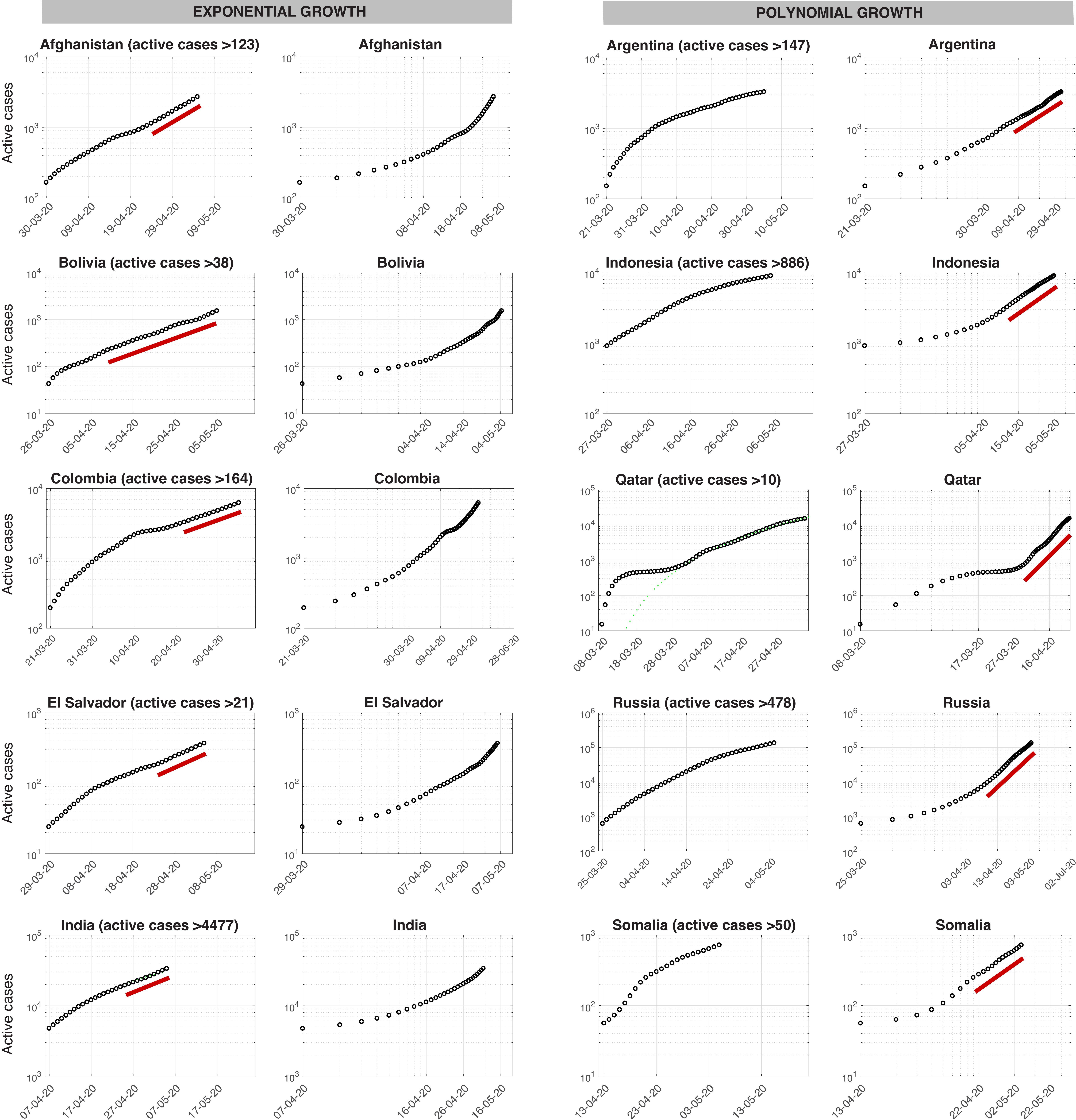}
\caption{Countries with the total number of active cases growing exponentially (two columns on left) or polynomially (two columns on right). Semilog plot (left) and double log plot (right) are displayed for each country. Straight lines indicate exponential and polynomial  growth in semi log and double log plots, respectively.  The data is shown from the epidemic onset until May  5, 2020 \cite{data}.
\label{figure1}}
\end{figure}

After EG we observe a systematic transition to polynomial growth (PG) in data during early epidemic stages. We support our claim here by Fig.~\ref{figure1} that shows the number of active cases in selected countries during the pandemic outbreak (the data source \cite{data}, reported data from May 5, 2020).  The figure displays particular illustrative cases, see Section \ref{s:Data} for a systematic survey of all observed countries. For each country displayed we show the time series of active cases both on the semilogarithmic and on the double logarithmic scales. On the semilogarithmic plot we detect a divergence from the initial exponential trend while on the double log plot a polynomial growth (represented as a linear trend)  is emerging after the initial EG. 

\subsection{Polynomial growth literature}
\label{s:PGE}
Polynomial growth in epidemic data and its eventual sources were previously discussed in literature, including a COVID-19 context.   
Ziff \& Ziff \cite{Ziff} observed PG in total number of deaths in China. Similarly,  Li \& Deng \cite{Li} study COVID-19 data in mainland China and observe an evidence of PG in the form $N(t) = At^{p}$ with exponents $p=2.48$, $p = 2.21$, and $p = 4.26$ in the total number of confirmed positive cases, confirmed deaths, and recovered cases, respectively. Szab\'o \cite{Szabo} in a theoretical work explains the PG in the total number of infected by the topology changes in the contact networks that has a major impact on the final extent of the pandemic, see also  \cite{Block} for a survey of impact of network topology changes on COVID-19 pandemic parameters. Szab\'o shows that the Barab\'asi-Albert  preferential attachment contact network with two connections of each new node and a simplified SIR dynamics implies total extent of the pandemic at only 4\% of the total population. That is a significant reduction from the levels expected from the SIR model (60-70\%) for the studied values  $2.2 < R_0 < 2.6$. The explanation of the lower value of an effective $R_0$ is that the most connected nodes are infected early in the epidemic and thus they effectively reduce the network connectivity. 
Manechein {\it et al.} \cite{Manchein2020} observed a polynomial growth of COVID-19 in all studied countries and identified a high degree of correlation between countries. 
Also, Maier \& Brockmann \cite{Maier} propose a modification of a SIR model that produces sub-exponential growth of infected individuals.  Due to the so-called containment strategies they introduce in their model loss terms for the number of susceptible and infected individuals that effectively force an exponential decay that should mimic (self-)isolation. 
Komarova \& Wodarz \cite{Komarova} also observe both EG and PG in COVID-19 data and show that a pandemic progress  in various countries is similar to Italy and can be mapped onto a universal timescale by accounting for the country-specific time delay. They suggest the PG is due to spatially structured population where the mixing between individuals is reduced. The metapopulation is modeled as a grid of patches with free mixing within each patch and contact with neighboring patches. No long-distance transmission is allowed in the model.
Brandenburg \cite{Brandenburg} analyzed COVID-19 data from China and observed that the number of fatalities and infections closely followed a quadratic law. The author provided an intuitive argument based on a local-spread model  through a periphery of the infected area that a quadratic rate of spread appears when local transmission in a regular 2D-lattice is the only mean of transmission. The polynomial spread by a fractal network of social interactions was also shown in \cite{Calvo} with the estimated polynomial exponent $p  \doteq 3.75$. See also \cite{BK2020,Radiom2020,Merrin} for surveys of demonstrating PG in COVID-19 pandemic data, growth rates estimates for individual countries, and suggested explanations of the PG regime.

\subsection{PG regime in SIR models}
\label{s:PG-SIR}
The basic SIR model is not consistent with a systematic approximate polynomial growth in the infected population. Otherwise, if $I(t) \approx At^p$, $p >0$, on a interval $t \in [t_1, t_2]$, then $dI/dt \approx (p/t) I$. 
Consequently from \eq{Ieq} 
$$
S(t) \approx \frac{N}{\beta} \left(\frac{p}{t} + \gamma\right) \qquad \qquad 
\mbox{and} \qquad \qquad
\frac{dS}{dt} \approx - \frac{N}{\beta} \, \frac{p}{t^2}\, .
$$
From \eq{Seq} it follows that
\begin{equation}
t^p \left( \frac{1}{t} + C\right) \approx t^{-2}\, .
\label{inconsistent}
\end{equation}
However, the approximation \eq{inconsistent} on the interval $t \in [t_1, t_2]$ of a significant length is inconsistent with the assumption $p >0$.
Note that modifications and extensions of SIR model can eventually agree with the observed polynomial growth phase in the infected population, however, we do not explore this question here.

\section{Polynomial growth with exponential decay regime (PGED)}
\label{s:PGED}
During late epidemic stages  in individual countries we  systematically observe a transition to a universal scaling form (ansatz) for the number of active cases in all considered countries. 
In this phase the epidemic wave reaches its peak after which the number of active cases decays. 
The scaling  has the form
\begin{equation}
I(t) = \frac{A}{T_G} \left( \frac{t}{T_G} \right)^\alpha \exp \left( -\frac{t}{T_G} \right) \, .\label{eq:Vazquez}
\end{equation}
Here $A, T_G$ and $\alpha$ are the model parameters. The scaling \eq{eq:Vazquez} is a combination of a polynomial growth factor $(t/T_G)^{\alpha}$ with an exponential decay $\exp(-t/T_G)$.  Therefore we refer to \eq{eq:Vazquez} as polynomial growth with exponential decay (PGED). It was derived
 for the size of the infected population by  Vazquez \cite{Vazquez} 
who used a branching process to describe the epidemic dynamics in a population interacting on a scale-free contact network. Ziff \& Ziff \cite{Ziff} use the PGED scaling on reported COVID-19 data and  claim that public measures and social distancing enforced yield a fractal type contact network on which the epidemic transmission is strongly limited by the network topology. 
We note that the polynomial growth models in the literature discussed  in Section~\ref{s:PGE} can be also adopted to account for the exponential decay factor as well by an inclusion of a constant rate of loss from the infected population (similarly as in \cite{Maier}). 

We  use \eq{eq:Vazquez} to match the observed pandemic data, particularly the number of  active infection cases as reported in \cite{data}. Note again that no prefactors are used here so the model only describes the dynamics of the reported data. 
The function $I = I(t)$ in \eq{eq:Vazquez} has a maximum at  $t = T_M=\alpha T_G$ where it reaches the value  $P = A(\alpha/e)^{-\alpha} T_G^{-1}$.
Note that the inflection points of the function $I = I(t)$ are located at $T_I^{\pm} = \left(\alpha \pm \sqrt{\alpha}\right)T_G$, particularly the time $t = T_I^-$ plays an important role in the observed epidemic data as it corresponds to a moment at which the growth of the number of active cases reaches its maximum and starts to decrease. 
In the SIR model the ratio of the infected population at the inflection point (during the growth phase) and at the point of maximum is equal to 1/2. However, in \eq{eq:Vazquez} this ratio is given by $\left(\frac{\alpha - \sqrt{\alpha}}{\alpha}\right)^{\alpha} e^{\sqrt{\alpha}}$. This expression is equal to $1/2$ for $\alpha \doteq 6.23$. For many countries we observe  $\alpha < 6.23$  and thus the slowdown of the epidemiological curve (inflection point) occurs at (often at a significantly) smaller fraction of the population than predicted by the SIR model.

An interpretation of the parameters $A, \alpha$ and $T_G$ is not completely straightforward and thus we reparametrize the model by a parameter combinations that correspond to naturally observed quantities. The equation  \eq{eq:Vazquez} rewritten using the parameters $P, T_M$ and $\alpha$ has the form
\begin{equation}
I(t) = P \left( \frac{t}{T_M} \right)^\alpha \exp \left[ \alpha \left( 1-\frac{t}{T_M} \right) \right]\, . \label{eq:Vazquez2}
\end{equation}

\subsection{Data fitting --- Methods}
\label{s:DF}
The model parameters in \eq{eq:Vazquez2} were inferred by nonlinear least squares optimization in MATLAB{\textcopyright}, see Tables~\ref{tab1}\ref{tab2} for the values obtained for individual countries. Polynomial and exponential decay factors in \eq{eq:Vazquez} motivate to use logarithmically rescaled data ($\log I(t)$ or both $\log I(t)$ and $\log t$) in optimization. However, we use non-rescaled values instead as the PGED trend is present only in later phases of the epidemic, particularly after the implementation of mitigation measures and social distancing (with a delay for its manifestation in the reported data). Using non-rescaled data allows us to globally fit the whole data set as the early epidemic data has only a small weight in the optimization due to its relative magnitude unlike in the case of the rescaled data. The fit thus does not require any prior (or fitting) for the time of transition to PGED.%
\footnote{We survey all countries systematically except for Mauritius where we have discarded early epidemic data before fitting the parameters of the model. These non-rescaled data points skew the fitted distribution significantly due to an unusually large weight. After the discard of the data, the fit of the recent data for Mauritius is cmparable with other countries. The nonstandard trend in data may be a consequence of extremely strict measures applied very early in Mauritius. We also have set the origin of the data for Singapore to later dates as the current data show already a second wave of pandemic.}

In general, fitting polynomial growth to data is very sensitive to a choice of the origin of the fitted time series. Therefore we have set the data starting point in all considered countries systematically. To eliminate the effect of stochasticity in small data we have disregarded all the data points in time series before the infected population in country reached a set threshold $N_0$. For Italy we set the threshold to 200 while for all other countries the threshold was normalized --- proportionally increased or decreased in agreement with the ratio of population size of the studied country to population of Italy (with minimum threshold set to 10 for countries with a small population).  

To eliminate obvious irregularities in daily reporting we smooth the studied data.
The irregularities appear naturally as the testing procedures and protocols impose systematic nonuniformity: populations in large clusters are discovered simultaneously, there is a systematic delay in contact-tracing testing, limited testing capacity, batch testing, etc.
We use linear smoothing on the increments and decrements of the number of active cases via moving averages through seven days (weights: $(1,3,6,7,6,3,1)/27$) that corresponds to three iterations of local averaging of three consecutive days. %
\footnote{There are significant irregularities in reporting of the number of recovered cases in some countries. United Kingdom does not report recovered cases hence we have removed it from our analysis. Some other countries only report the recovered cases weekly or they report them only very seldom. Thus for  Finland, Greece, Ireland, and Mexico we have used the simple equal weight moving average over 20 days for the recovered cases data. In multiple other irregularly reporting countries such smoothing has a small effect on the active cases model parameters and thus we did not apply it.}

\subsection{PGED regime in SIR Models}
\label{s:PGsection}
Polynomial growth is not consistent with the SIR model for an extented period. However, we consider very instructive and useful to reconcile the SIR models with the PGED regime as such a reconciliation would allow to build PGED directly into the SIR type models. Therefore we study whether the form (equivalent to \eq{eq:Vazquez2})
\begin{equation}
I(t) \approx B t^{\alpha} e^{-t/T_G}
\label{PGED}
\end{equation} 
can solve a  SIR type model. Equation \eq{PGED} implies 
\begin{equation}
\frac{dI}{dt} = \left(\frac{\alpha}{t} - \frac{1}{T_G}\right) I \, .
\label{PGder}
\end{equation}
We can now compare \eq{PGder} with \eq{Ieq} and identify 
$$
\gamma = \frac{1}{T_G}\qquad \qquad \mbox{and} \qquad \qquad 
\beta \frac{S}{N} =  \frac{\alpha}{t}\, .
$$
Thus, the exponential decay $e^{-t/T_G}$ term in \eq{PGED} can be seen as a consequence of the infected removal rate $\gamma = 1/T_G$ in the SIR model. 
Furthermore, the term $\beta S / N$ corresponds to $\alpha/t$.
Finally, we express this dependence in terms of the relative reproduction number $R_0$:
$$
R_0(t)  = \frac{\beta}{\gamma} \, \frac{S}{N} = \frac{T_M}{t}\, ,
$$
where $T_M = \alpha T_G = \alpha / \gamma$ is a constant, see Section~\ref{s:PGED} for its interpretation as the time of the epidemic peak. Therefore PGED implies an algebraic decay of $R_0$ in the SIR model.

Note that this result is in agreement with the argument about the reduction of infection transmissibility in \cite{He} and also with the $R_0$ analysis in 
study \cite{Flaxman} of the impact of non-pharmaceutical interventions in European countries. There the change of the reproduction number $R_0$ is modeled as an average percentage reduction per specific type of intervention and the analysis is based on a Bayesian approach using data on number of infected and number of deaths, which are more reliable.
Also note that as $t \rightarrow 0^{+}$ the relative reproduction number diverges to $\infty$. This is in an agreement with the model of \cite{Szabo} where the fat-tail power-law distribution of $R_0$ in the population has initially an infinite mean that  with the epidemic outbreak reduces to finite values. Average $R_0$ further reduces during the epidemic by a gradual elimination of the individuals with a high values of $R_0$ from the susceptible population --- the individuals with high values of $R_0$ are easily infected and  removed from the susceptible class as the first. 

\section{Three Phases of Epidemics --- Summary}
\label{s:3P}
Based on the presented analysis we expect three phases of a single epidemic wave.

\begin{itemize}
\item
{\bf Initial Exponential Phase.}
During the initial phase (with no applied mitigation measures) we expect the exponential growth of the infected population $I = I(t)$ as discussed in Section~\ref{s:EG}: 
\begin{equation*}
I(t) \approx I_0 \exp\Big(\gamma(R_{0}(0)-1)t\Big)\, .
\end{equation*}  
\item
{\bf Polynomial Growth Phase.}
After the initial phase we expect a short transient phase during which $I = I(t)$ smoothly transitions from the initial exponential phase to the polynomial growth phase described in Section~\ref{s:PG}. During this phase 
\begin{equation*}
I(t) \approx Ct^p\, .
\end{equation*}
\item
{\bf Final Polynomial Growth with Exponential Decay Phase.}
After an introduction of the mitigation measures and social distancing and a subsequent delay necessary for their appearance in the reported data we expect a transition of $I = I(t)$ from the PG phase to PGED phase described in Section~\ref{s:PGED}. During this phase
\begin{equation*}
I(t) \approx  P \left( \frac{t}{T_M} \right)^\alpha \exp \left[ \alpha \left( 1-\frac{t}{T_M} \right) \right]\,
\end{equation*}
\end{itemize}
Note that is some countries the mitigation measures were applied when the number of infected was low or zero (Slovakia, Lithuania) and the initial exponential phase was too short to be identified in the data.  
Also note that during the PG phase the function $I = I(t)$ is convex ($p >1$ for all observed countries). For such a function to reach its local maximum it must first go through an inflection point. Unless the inflection point appears during the transient phase between PG and PGED the PG phase must connect to the PGED phase before the PGED phase reaches its inflection point, i.e., before the time $t = T_I^-$. This phenomenon was also observed in data of all surveyed countries and it helps to improve predictions for countries  that have not reached the PGED phase yet.

%

\section{Data Analysis Results}
\label{s:Data}
We conducted a systematic survey of COVID-19 pandemic data (\cite{data}, the  last reporting day May 9, 2020) for all countries where the time series are sufficiently long to display a consistent trend (in total 118 countries). In each country (together with all its territories) we consider the number of active cases equal to the total number of reported confirmed infections decreased by the sum of reported total number of recovered and deaths. For a characterization of the epidemic progression we use the following phases: the initial exponential phase (EG), the polynomial growth phase (PG), and 
the polynomial growth with exponential decay phase (PGED). The final PGED phase has two checkpoints, the  inflection point (I) and the epidemic peak---the point of maximum of the active infected population (M), after which the number of infected consistently decreases (D).

\subsection{Classification of Individual Countries}
\begin{figure}
\centering
\includegraphics[width=0.75\textwidth]{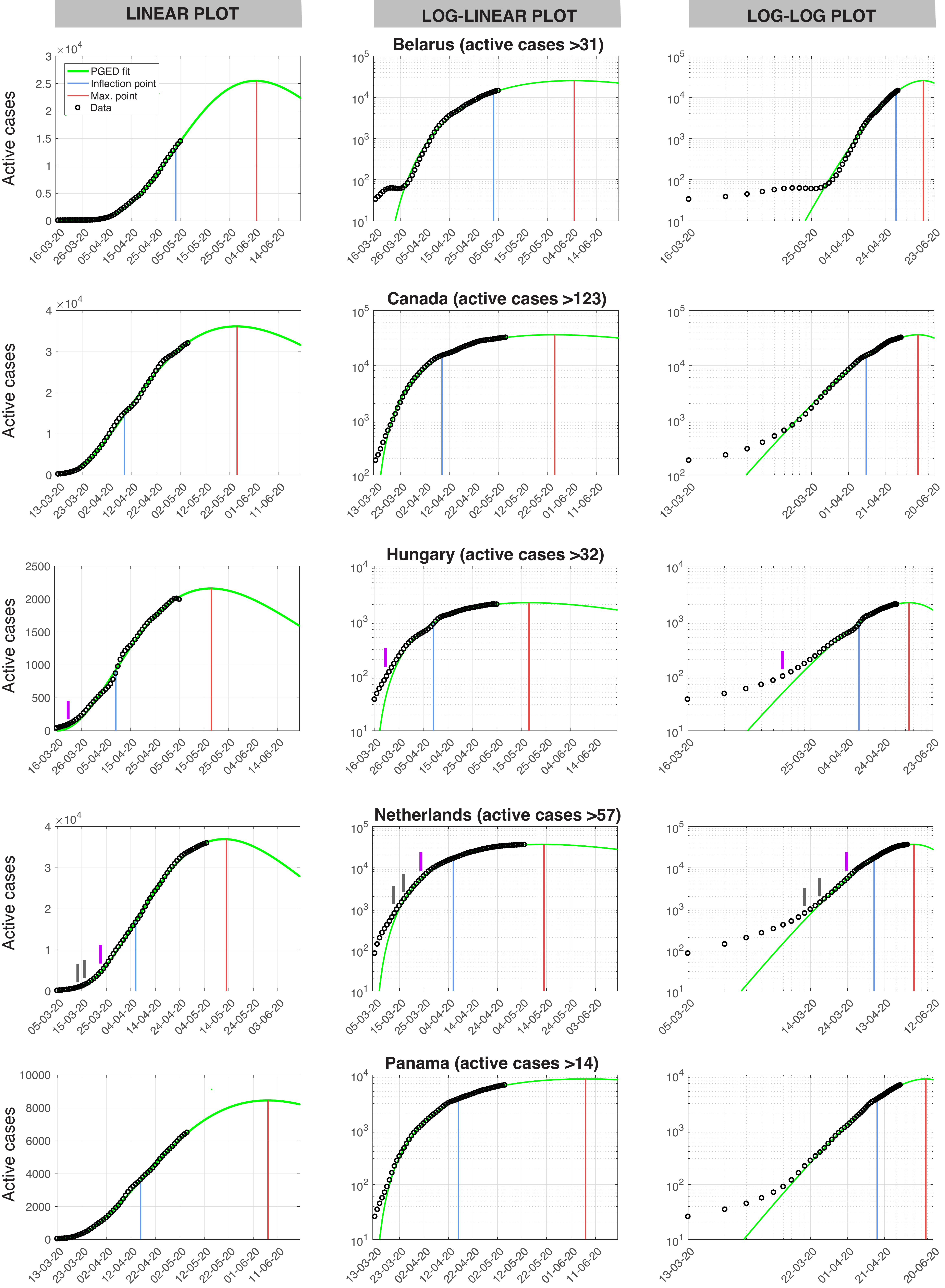}
\caption{Selected countries beyond their inflection point of PGED regime (part I). The country-specific locations of the PGED inflection point (blue) and the epidemic peak (red) are marked by vertical lines. For model parameters see Table~\ref{tab2}. Dates of applicaton of mitigation measures are indicated for some countries (data \cite{Flaxman,data_measures}) by the gray pointers. The magenta pointers indicate the time 10 days after the application of the first measure. The best PGED model fit is plotted in green. 
Linear (left), semilogarithmic  (center), and double log plot (right) are shown for each country. Straight lines indicate EG and PG in semi log and double log plots, respectively.  The data is shown from the epidemic onset until May  5, 2020 \cite{data}.
\label{figure2}}
\end{figure}

\begin{figure}
\centering
\includegraphics[width=0.75\textwidth]{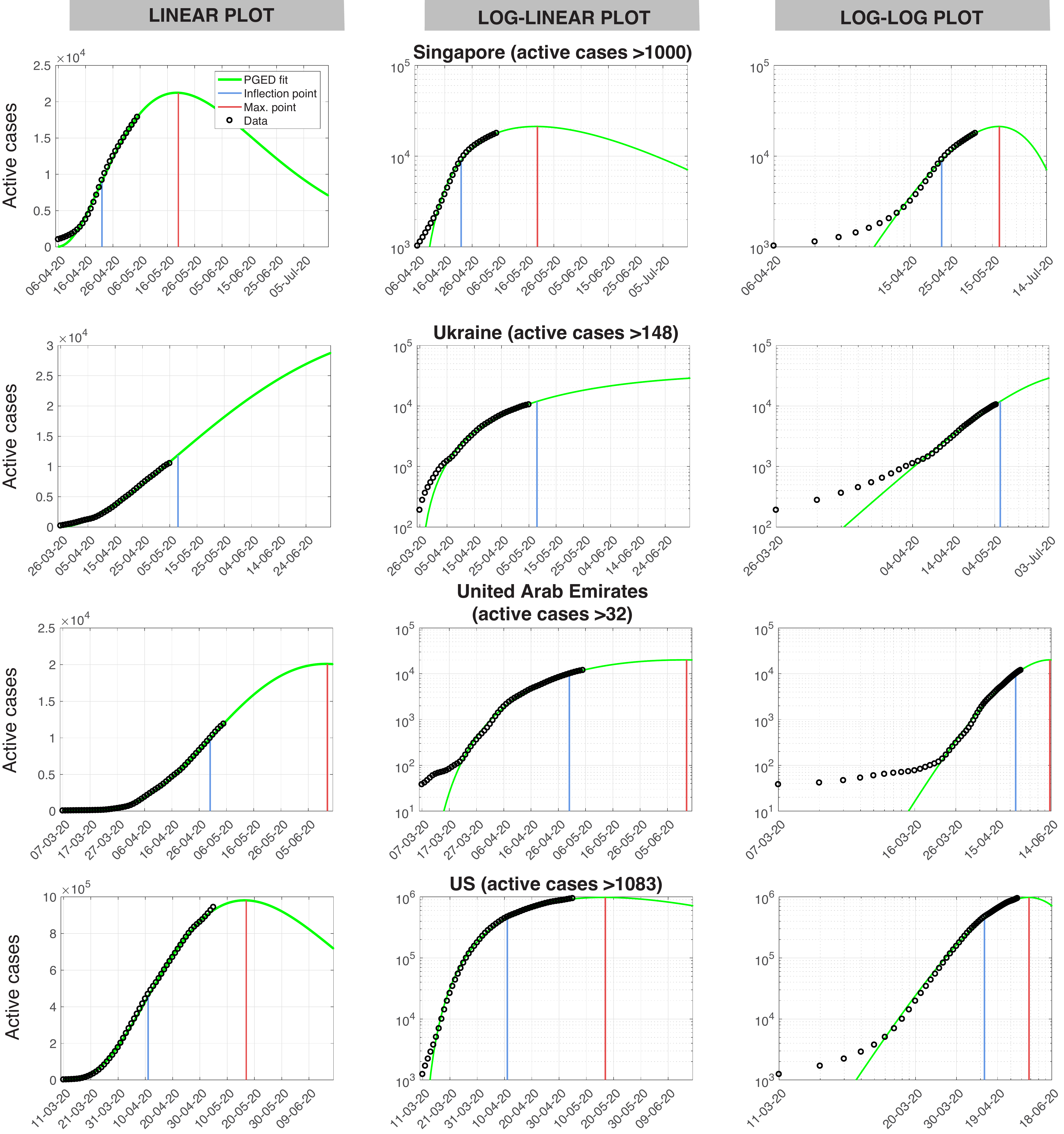}
\caption{
Selected countries beyond their inflection point of PGED regime (part II). For model parameters see Table~\ref{tab2}. The data is shown from the epidemic onset until May  5, 2020 \cite{data}.
\label{figure3}}
\end{figure}

\begin{figure}
\centering
\includegraphics[width=0.75\textwidth]{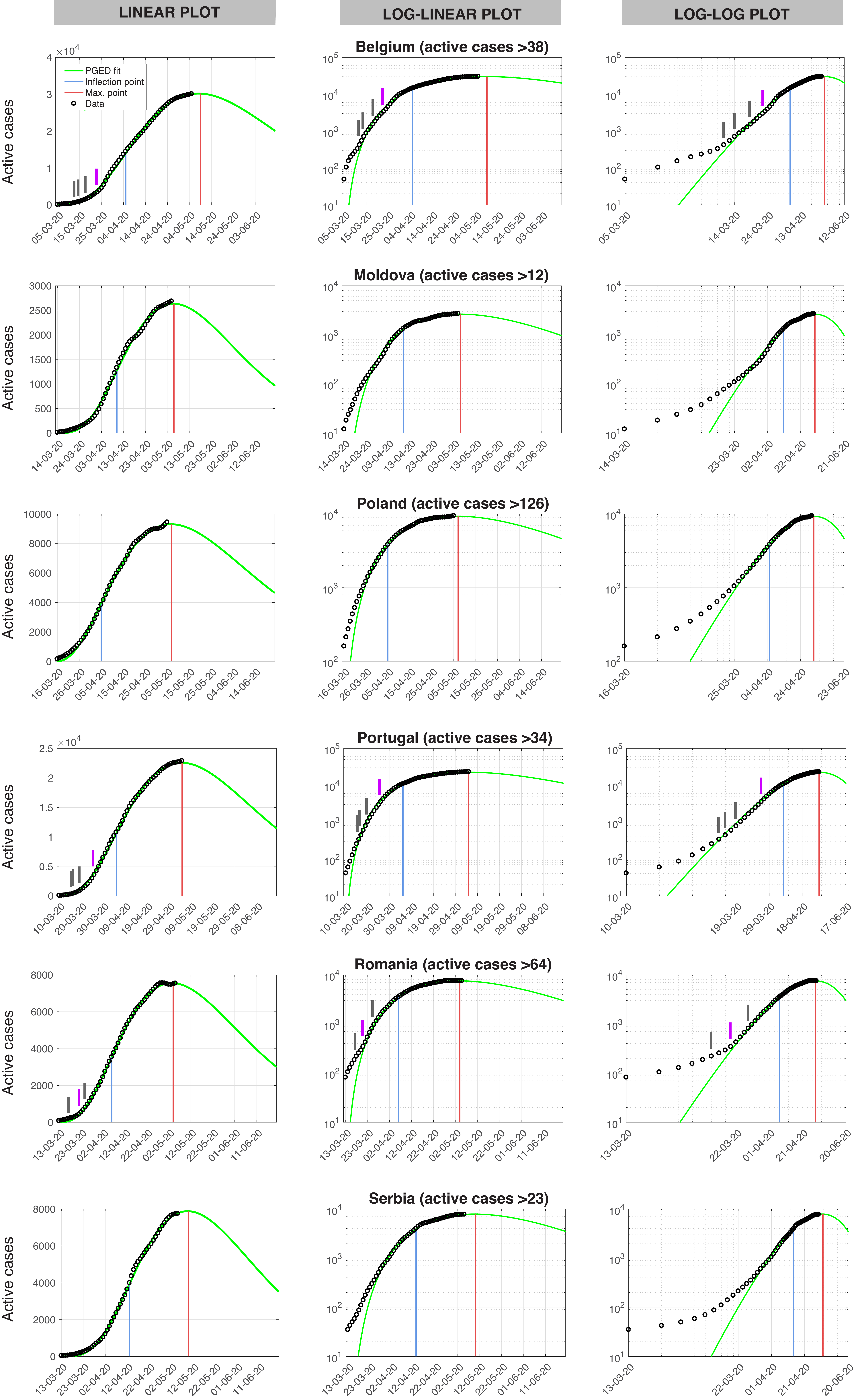}
\caption{Selected countries close to their epidemic peak. The peak is defined  as the maximal total number of active cases. The best PGED model fit is plotted in green. For model parameters see Tables~\ref{tab1}--\ref{tab2}. The data is shown from the epidemic onset until May  5, 2020 \cite{data}.
\label{figure4}}
\end{figure}

\begin{figure}
\centering
\includegraphics[width=0.75\textwidth]{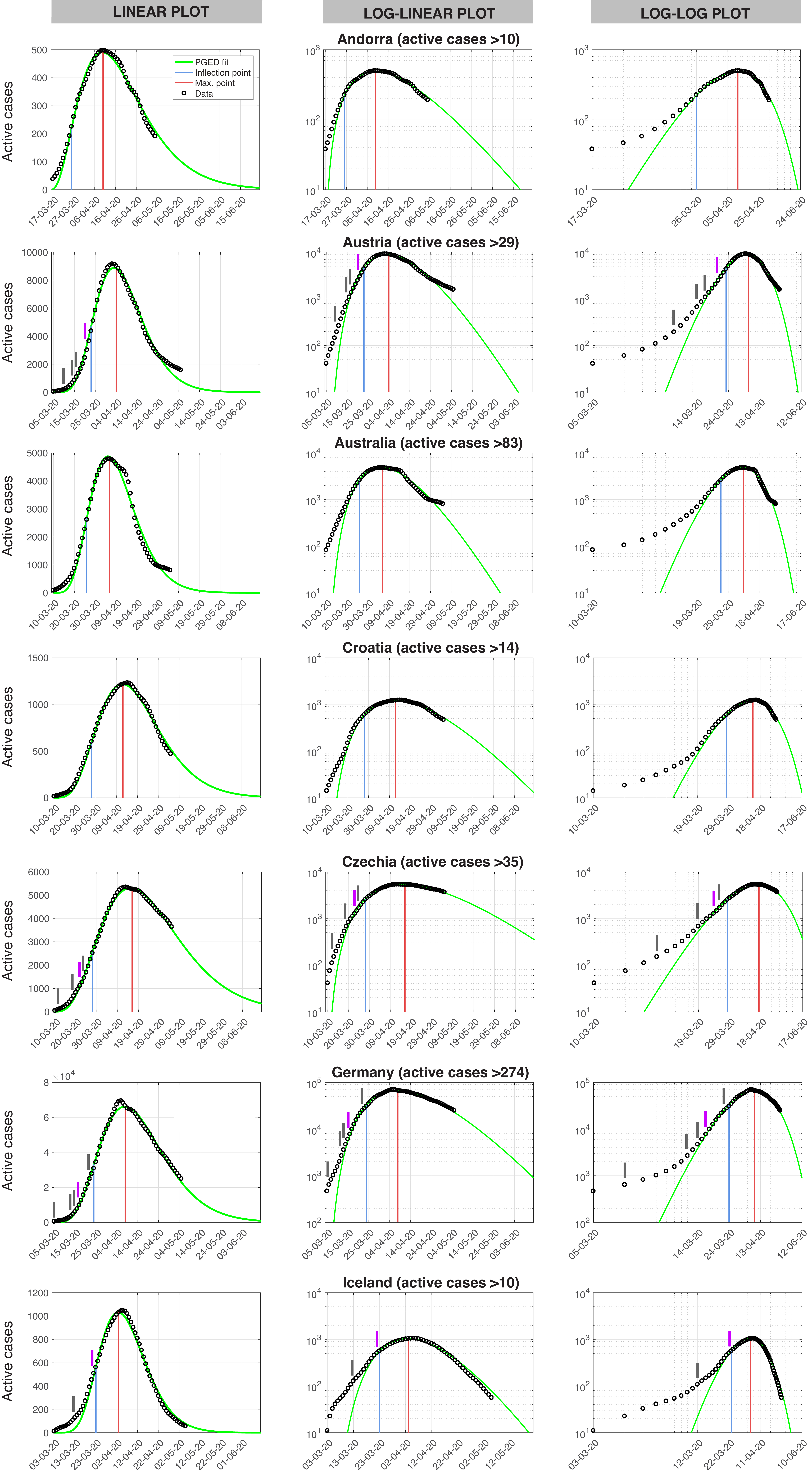}
\caption{Seected countries past their epidemic peak, part I. For model parameters see Table~\ref{tab1}.  
The data is shown from the epidemic onset until May  5, 2020 \cite{data}.\label{figure5}}
\end{figure}

\begin{figure}
\centering
\includegraphics[width=0.75\textwidth]{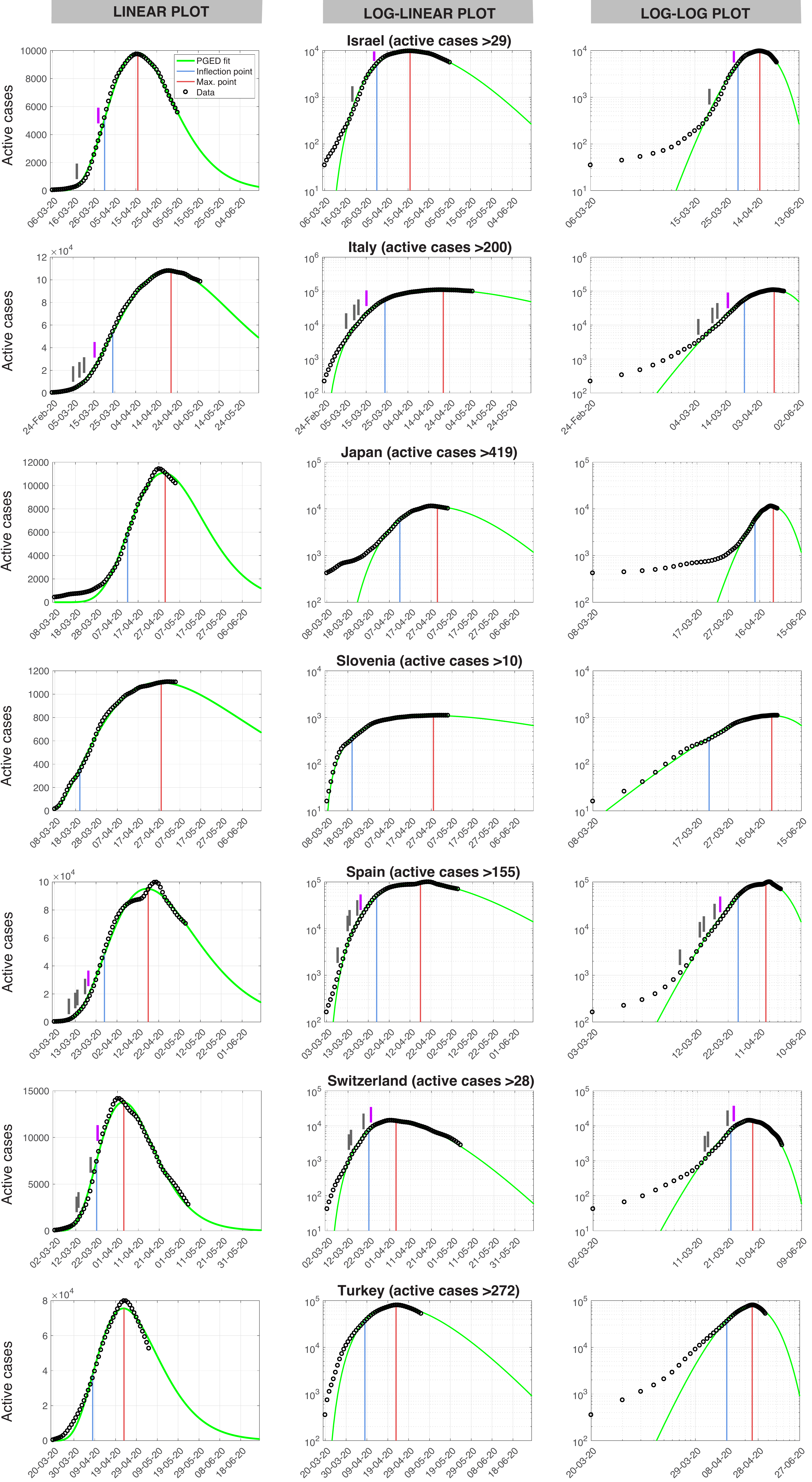}
\caption{Seected countries past their epidemic peak, part II. For model parameters see Tables~\ref{tab1}--\ref{tab2}.  
The data is shown from the epidemic onset until May  5, 2020 \cite{data}.
\label{figure6}}
\end{figure}

\begin{figure}
\centering
\includegraphics[width=0.75\textwidth]{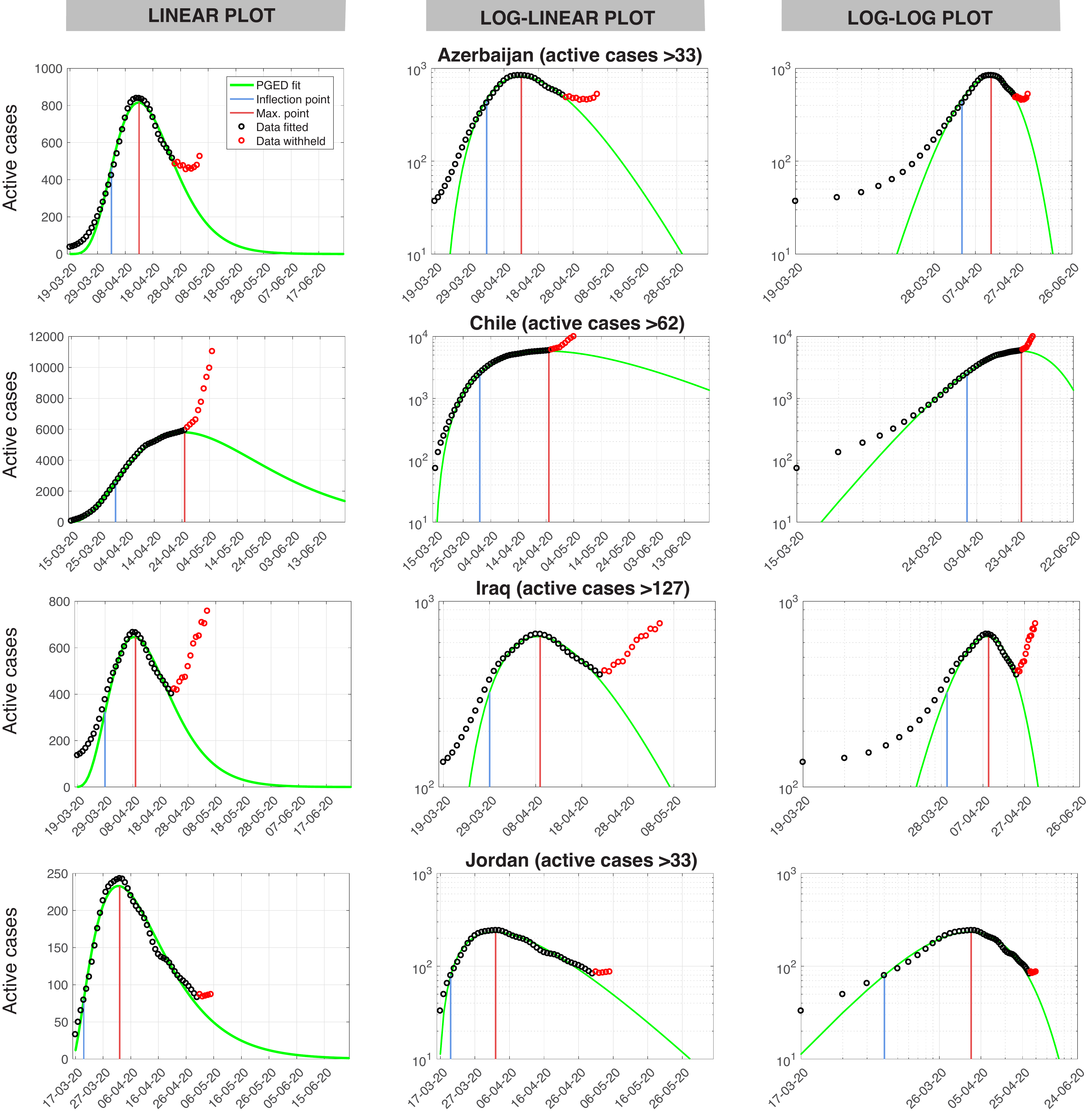}
\caption{Selected countries that are most likely entering into the second epidemic wave.  For model parameters see Table~\ref{tab2}.  
The data is shown from the epidemic onset until May  5, 2020 \cite{data}.
\label{figure7}}
\end{figure}

\begin{figure}
\centering
\includegraphics[width=\textwidth]{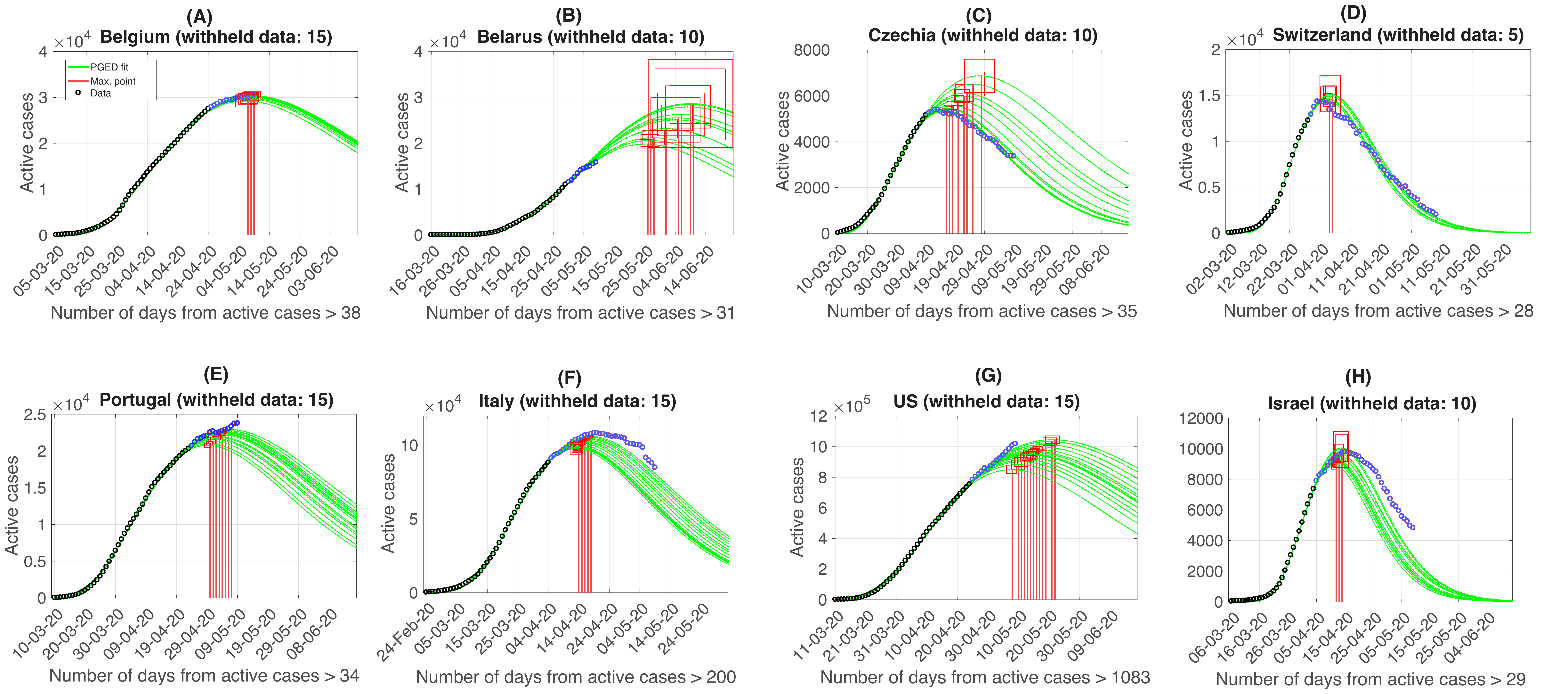}
\caption{Predictions based on PGED model. For each country we remove the last $n$ data points or last $n$ data points before the epidemic peak (if already reached), $0\le n \le D$  (see plot labels for the values of $D$) while always keeping the data points in black. PGED model parameters are inferred for each $n$. 95\% confidence intervals for two PGED parameters (time and population of the epidemic peak) inferred by nonlinear regression are shown as bounding boxes around the mean in red. 
The presented data display: small uncertainty, small overlapping confidence intervals (A), large uncertainty, not enough data (B), monotonicity, additional data shifts the peak earlier (C), 
well predicted location of the peak and the data past (D), monotonicity, additional data shifts peak later (E, F, G), well predicted location of the peak but data past the peak not well captured (H).The data is shown from the epidemic onset to May 9, 2020 \cite{data}.
\label{figure8}}
\end{figure}

Our results are presented in Tables~\ref{tab1}--\ref{tab3} that summarize the stage of the epidemic in all surveyed countries. If a country reached the PGED regime we report the estimated values of the related PGED parameters.   See Section~\ref{s:DF} for the methodology and remarks on exceptions made for some individual countries. 
We support our results in Fig.~\ref{figure1}--\ref{figure8} that display data from selected countries. The figures show total active cases data time series for countries that are organized by the epidemic phase. For each country presented we show the data on both linear and semilogarithmic plot (countries in EG and PG phase) and also on double logarithmic plot (countries in PGED phase). For the countries in the PGED regime we also show the best PGED fit to the data.%
\footnote{For technical reasons the plots correspond to data \cite{data} with the last data point on May 5, 2020.
Table~\ref{tab4} contains the estimated PGED parameters for the data shown in these figures.}

\begin{table}
\begin{center}
\begin{tabular}{@{}l c c c r r r r r}
Country&\phantom{I}I\phantom{I} &M &D &$T_M$\ \ \ &$P$\ 	\ \ &$\alpha$\ \ \ \	&$T_G$\ \ \ &$N_0$\ \\
\hline
Andorra & $\ast$&$\ast$&$\ast$&10-Apr & 492 & 2.6262 & 9.3431 & 10\\
Australia & $\ast$&$\ast$&$\ast$&05-Apr & 4848 & 6.1804 & 4.4026 & 83\\
Austria & $\ast$&$\ast$&$\ast$&03-Apr & 8836 & 6.8140 & 4.4141 & 29\\
Barbados & $\ast$&$\ast$&$\ast$&10-Apr & 54 & 1.5579 & 12.9616 & 10\\
Bosnia and Herz. & $\ast$&$\ast$&$\ast$&01-May & 942 & 2.4445 & 20.4171 & 11\\
Brunei & $\ast$&$\ast$&$\ast$&01-Apr & 95 & 1.9501 & 7.2960 & 10\\
Cambodia & $\ast$&$\ast$&$\ast$&13-Apr & 89 & 1.3276 & 8.0132 & 10\\
China & $\ast$&$\ast$&$\ast$&04-Apr & 55120 & 3.2582 & 6.4216 & 4609\\
Costa Rica & $\ast$&$\ast$&$\ast$&14-Apr & 527 & 3.3314 & 10.1959 & 17\\
Croatia & $\ast$&$\ast$&$\ast$&12-Apr & 1221 & 5.2767 & 6.3767 & 14\\
Cuba & $\ast$&$\ast$&$\ast$&24-Apr & 787 & 3.3270 & 9.9317 & 38\\
Cyprus & $\ast$&$\ast$&$\ast$&22-Apr & 662 & 4.1410 & 9.7971 & 10\\
Czechia & $\ast$&$\ast$&$\ast$&16-Apr & 5283 & 3.7804 & 9.9790 & 35\\
Denmark & $\ast$&$\ast$&$\ast$&10-Apr & 3268 & 3.2980 & 11.0206 & 19\\
Eritrea & $\ast$&$\ast$&$\ast$&18-Apr & 36 & 1.2907 & 11.2583 & 10\\
Estonia & $\ast$&$\ast$&$\ast$&24-Apr & 1346 & 2.5781 & 17.9813 & 10\\
Finland$^\ast$ & $\ast$&$\ast$&$\ast$&19-Apr & 2159 & 3.2336 & 13.2385 & 18\\  
France & $\ast$&$\ast$&$\ast$&30-Apr & 97250 & 4.7348 & 12.1991 & 222\\
Georgia & $\ast$&$\ast$&$\ast$&07-May & 346 & 3.8800 & 15.8168 & 12\\
Germany & $\ast$&$\ast$&$\ast$&08-Apr & 65981 & 5.2108 & 6.7248 & 274\\
Greece$^\ast$ & $\ast$&$\ast$&$\ast$&15-Apr & 1778 & 3.8199 & 10.8249 & 36\\
Iceland & $\ast$&$\ast$&$\ast$&03-Apr & 1034 & 8.2338 & 3.8499 & 10\\
Ireland$^\ast$ & $\ast$&$\ast$&$\ast$&18-Apr & 10287 & 10.3295 & 4.3819 & 16\\
Israel & $\ast$&$\ast$&$\ast$&16-Apr & 9831 & 6.7152 & 6.2335 & 29\\
Italy & $\ast$&$\ast$&$\ast$&20-Apr & 108081 & 4.3420 & 13.0521 & 200\\
Japan & $\ast$&$\ast$&$\ast$&28-Apr & 11004 & 10.3488 & 5.0056 & 419\\
Korea & $\ast$&$\ast$&$\ast$&14-Mar & 6996 & 1.9963 & 11.5002 & 171\\
Kosovo & $\ast$&$\ast$&$\ast$&24-Apr & 513 & 6.3367 & 4.7844 & 10\\
Latvia & $\ast$&$\ast$&$\ast$&17-Apr & 614 & 2.5766 & 13.8421 & 10\\
Lithuania & $\ast$&$\ast$&$\ast$&17-Apr & 979 & 3.1019 & 10.8176 & 10\\
Luxembourg & $\ast$&$\ast$&$\ast$&08-Apr & 2830 & 5.4882 & 5.1384 & 10\\
Malaysia & $\ast$&$\ast$&$\ast$&10-Apr & 2442 & 2.1384 & 14.8065 & 104\\
Malta & $\ast$&$\ast$&$\ast$&11-Apr & 323 & 5.9041 & 5.1157 & 10\\
Mauritius$^\ast$ & $\ast$&$\ast$&$\ast$&18-Apr & 281 & 9.8527 & 2.2119 & 10\\
Monaco & $\ast$&$\ast$&$\ast$&15-Apr & 80 & 2.7548 & 7.6893 & 10\\
Montenegro & $\ast$&$\ast$&$\ast$&11-Apr & 258 & 4.2175 & 5.5614 & 10\\
New Zealand & $\ast$&$\ast$&$\ast$&07-Apr & 879 & 4.4844 & 4.7651 & 16\\
N. Macedonia & $\ast$&$\ast$&$\ast$&18-Apr & 893 & 6.8064 & 5.4705 & 10\\
Norway & $\ast$&$\ast$&$\ast$&03-May & 7626 & 2.6264 & 24.3982 & 18\\
Panama & $\ast$&$\ast$&$\ast$&06-May & 5700 & 4.4716 & 12.3516 & 14\\
Portugal & $\ast$&$\ast$&$\ast$&07-May & 23169 & 3.2656 & 18.1356 & 34\\
Romania & $\ast$&$\ast$&$\ast$&03-May & 7537 & 3.5309 & 14.7572 & 64\\
Serbia & $\ast$&$\ast$&$\ast$&06-May & 7495 & 5.1916 & 10.5773 & 23\\
Slovakia & $\ast$&$\ast$&$\ast$&21-Apr & 877 & 4.1067 & 9.7964 & 18\\
Slovenia & $\ast$&$\ast$&$\ast$&29-Apr & 1109 & 1.7043 & 30.8777 & 10\\
\hline
\end{tabular}
\caption{Classification of countries, part 1. I -- past the inflection point, M -- around the maximum, D -- decreasing. Data \cite{data} from May 9, 2020.}
\label{tab1}
\end{center}
\end{table}

\begin{table}
\begin{center}
\begin{tabular}{@{}l c c c r r r r r}
Country&\phantom{I}I\phantom{I} &M &D &$T_M$\ \ \ &$P$\ 	\ \ &$\alpha$\ \ \ \	&$T_G$\ \ \ &$N_0$\ \\
\hline
Spain & $\ast$&$\ast$&$\ast$&17-Apr & 95129 & 4.5803 & 9.9396 & 155\\
Switzerland & $\ast$&$\ast$&$\ast$&04-Apr & 13770 & 6.1837 & 5.4694 & 28\\
Thailand & $\ast$&$\ast$&$\ast$&15-Apr & 1446 & 1.7718 & 8.3237 & 230\\
Trinidad and T. & $\ast$&$\ast$&$\ast$&11-Apr & 96 & 1.0373 & 14.6813 & 10\\
Tunisia & $\ast$&$\ast$&$\ast$&17-Apr & 717 & 2.3942 & 11.9326 & 38\\
Turkey & $\ast$&$\ast$&$\ast$&22-Apr & 75817 & 5.7098 & 6.0236 & 272\\
Uruguay & $\ast$&$\ast$&$\ast$&05-Apr & 276 & 1.0004 & 21.4012 & 11\\
Uzbekistan & $\ast$&$\ast$&$\ast$&20-Apr & 1236 & 3.8443 & 5.9905 & 109\\
Vietnam & $\ast$&$\ast$&$\ast$&01-Apr & 145 & 3.9264 & 7.1140 & 10\\ 
Azerbaijan & $\ast$&$\ast$&$\nearrow$&13-Apr & 816 & 5.6844 & 4.5501 & 33\\
Burkina Faso & $\ast$&$\ast$&$\nearrow$&07-Apr & 271 & 1.5322 & 10.9200 & 65\\
Chile & $\ast$&$\ast$&$\nearrow$&24-Apr & 5810 & 2.7035 & 15.2673 & 62\\
Djibouti & $\ast$&$\ast$&$\nearrow$&22-Apr & 738 & 14.8228 & 1.9548 & 10\\ 
Iran & $\ast$&$\ast$&$\nearrow$&06-Apr & 27960 & 5.9122 & 6.5415 & 271\\
Iraq & $\ast$&$\ast$&$\nearrow$&08-Apr & 612 & 2.2660 & 9.2597 & 127\\
Jordan & $\ast$&$\ast$&$\nearrow$&02-Apr & 231 & 1.5561 & 10.6755 & 33\\
Kyrgyzstan & $\ast$&$\ast$&$\nearrow$&18-Apr & 356 & 3.5563 & 7.3488 & 21\\
Lebanon & $\ast$&$\ast$&$\nearrow$&21-Apr & 549 & 1.9582 & 22.7921 & 23\\
Madagascar & $\ast$&$\ast$&$\nearrow$&11-Apr & 85 & 2.0474 & 9.6303 & 10\\
\hline
Belgium & $\ast$&$\ast$& &09-May & 30304 & 3.7866 & 17.3035 & 38\\ 
Eq. Guinea & $\ast$& $\ast$      &      &13-May & 424 & 8.6780 & 5.6666 & 10\\ 
Hungary & $\ast$&$\ast$& &11-May & 1998 & 2.9094 & 19.5773 & 32\\  
Jamaica & $\ast$&$\ast$& &08-May & 422 & 10.7664 & 4.8224 & 10\\ 
Moldova & $\ast$&$\ast$& &08-May & 2714 & 3.9338 & 14.2948 & 12\\ 
Morocco & $\ast$&$\ast$& &08-May & 3349 & 2.9033 & 16.1095 & 119\\ 
Netherlands & $\ast$&  $\ast$    &      &12-May & 37007 & 3.5965 & 19.2411 & 57\\ 
Poland & $\ast$&$\ast$& &08-May & 9374 & 2.6627 & 20.1358 & 126\\ 
\hline
Singapore$^\ast$ & $\ast$&      & $\nearrow$ &23-May & 22770 & 2.3568 & 20.4616 & 1000$^\ast$\\
Belarus & $\ast$&      &      &25-May & 19714 & 6.8194 & 10.3855 & 31\\
Canada & $\ast$&      &      &21-May & 34740 & 2.7003 & 25.9996 & 123\\
Ecuador & $\ast$&      &      &13-May & 26778 & 6.9447 & 8.2141 & 57\\
Eswatini & $\ast$&      &      &15-Jul & 341 & 1.3660 & 65.4323 & 10\\
Maldives & $\ast$&      &      &19-May & 922 & 6.2324 & 6.7453 & 10\\
Saudi Arabia & $\ast$&      &      &26-May & 34075 & 6.7335 & 10.6732 & 112\\
Sweden & $\ast$&      &      &20-May & 18932 & 3.8054 & 20.5556 & 34\\
Ukraine & $\ast$&      &      &18-Jun & 17397 & 2.3863 & 35.8039 & 148\\
UAE & $\ast$&      &      &03-Jun & 17149 & 5.6187 & 15.7821 & 32\\
USA & $\ast$&      &      &22-May & 1061497 & 3.2211 & 22.8083 & 1083\\
Oman & $\ast$&      &      &14-May & 2147 & 5.1634 & 11.2438 & 16\\
\hline
\end{tabular}
\caption{Classification of countries, part 2.  
 I -- past the inflection point, M -- around the maximum, D -- decreasing. 
 The symbol $\nearrow$ indicates apparent emergence of a next epidemic wave; in that case 
the parameters were inferred disregarding the most recent data attributed  to the second wave. 
Data \cite{data} from May 9, 2020}
\label{tab2}
\end{center}
\end{table}

\begin{table}
\begin{center}
\begin{tabular}{@{}l c c l c c}
Country&EG &PG &Country&EG &PG \\ \hline
Afghanistan &       &  $\ast$ &Russia  &       &  $\ast$\\
Bahrain  &       &  $\ast$&San Marino  &       &  $\ast$\\
Bangladesh &       &  $\ast$&Senegal  &       &  $\ast$\\
Benin &       &  $\ast$& Somalia  &       &  $\ast$\\
Bolivia  &       &  $\ast$&Chile &       &  $\ast$\\
Bulgaria  &       &  $\ast$&Brazil &   $\ast$    &  ?\\
Chad  &       &  $\ast$&Central African Republic & $\ast$& \\
Colombia  &       &  $\ast$&Cote d'Ivoire & $\ast$& \\
Dominican Republic  &       &  $\ast$&Egypt & $\ast$& \\
Guatemala  &       &  $\ast$&El Salvador & $\ast$& \\
Indonesia  &       &  $\ast$&Gabon & $\ast$& \\
Kazakhstan  &       &  $\ast$&Ghana & $\ast$& \\
Kuwait  &       &  $\ast$&India & $\ast$& \\
Mexico$^\ast$  &       &  $\ast$&Nigeria & $\ast$& \\
Pakistan  &       &  $\ast$&Paraguay & $\ast$& \\
Peru  &       &  $\ast$&South Africa & $\ast$& \\
Philippines  &       &  $\ast$&Sudan & $\ast$& \\
Qatar  &       &  $\ast$\\ \hline
\end{tabular}
\caption{Classification of countries, part 3.  E -- exponential phase, PG -- polynomial phase. Data \cite{data} from May 9, 2020}
\label{tab3}
\end{center}
\end{table}

\begin{table}
\begin{center}
\begin{tabular}{@{}l c c c r r r r r}
Country&$T_M$&$P$&$\alpha$&$T_G$&$A$&$N_0$\\
\hline
Australia		&07-Apr-20&4,827&5.0672&5.642&1,160&83\\
Austria		&03-Apr-20&9,145&7.8921&3.7765&7.6746&29\\
Germany	&07-Apr-20&67,158&5.8838&5.8429&4,173&274\\
Iceland		&04-Apr-20&1,011&6.7469&4.8172&10.56&10\\
Jordan		&01-Apr-20&240&1.8785&8.5951&4,124&33\\
Korea		&14-Mar-20&7,278&2.3738&9.5062&95,439&171\\
New Zealand	&08-Apr-20&894&4.4955&4.8311&450.1&16\\
Switzerland	&03-Apr-20&14,125&7.3073&4.5191&46.43&28\\  \hline
Cameroon	 &12-Apr-20&703&2.3132&7.1547&7303&83\\ 
Croatia		 &14-Apr-20	&1,232&4.1544&8.7288&1,846&14\\
Czechia		 &17-Apr-20	&5,464&3.8142&10.119&15,188&35\\
Israel		 &13-Apr-20	&9,589&8.2554&4.7673&4.756&29\\
Italy			 &16-Apr-20&103,336&4.8895&10.8482&63,523&200\\
Lithuania	 &15-Apr-20&952&3.3056&9.6762&4,825&10\\
Malaysia		 &09-Apr-20&2,516&2.4584&12.5819&40,522&104\\
Spain		 &11-Apr-20&88,035&	6.0890	&6.5561	&4,252	&155\\ \hline
Belgium		&25-Apr-20&23,499&	4.9788	&10.3694	&1,947	&38\\
Bulgaria		& 12-May-20&761&1.4078&42.7808&2,223&23\\
Canada		&24-Apr-20&20,134&	3.7617	&11.4505	&7,922	&123\\
Chile		&22-Apr-20&5,644&	2.8983	&13.5026	&3,285	&62\\
Greece		&26-Apr-20&1953&	2.5529	&20.3868	&6,745	&35\\
Latvia		&20-Apr-20&652&	2.3804	&16.425	&14,679	&10\\
Netherlands	&13-May-20&36,901&	3.4973	&20.0621	&306,680	&57\\
Norway		&20-Apr-20&6,739&	3.5642	&14.4041	&36,056	&18\\
Portugal		&26-Apr-20&19,675&	4.2362	&11.4094	&34,266	&34\\
Romania		&08-May-20&8,322&	3.2727	&17.4427		&79,068	&64\\
Slovenia		&24-Apr-20&1,077&	1.8851	&25.5327		&54,809	&10\\
US 			&29-Apr-20&703,430&4.699  &10,6447	&571,928	&1083\\
\hline
\end{tabular}
\caption{Table of the PGED parameters used for Fig. 1--8 in the main article. The parameters were inferred from the dataset [23]  containing time series up to May 5, 2020.}
\label{tab4}
\end{center}
\end{table}

\subsection{Countries in EG and PG phases (05-May-2020)} 
In Figure~\ref{figure1} we present two groups of countries in the early stages of the epidemic: the countries in the EG phase (Afghanistan, Bolivia, Colombia, El Salvador, India) and countries in the PG phase (Argentina, Indonesia, Qatar, Russia, Somalia). To demonstrate an evidence of EG and PG, respectively,  
we compare the recent data with a straight line in the respective plot. 

\subsection{Countries in the PGED phase past inflection point (05-May-2020)}
Figures~\ref{figure2},~\ref{figure3} provide an graphical overview of selected countries past their inflection point in PGED regime but still ahead of their epidemic peak (both times for the fitted PGED approximation are indicated in the corresponding figure).  For selected countries we indicate the date \cite{Flaxman,data_measures} of the most severe social mitigation measures applied in the displayed country  and also a date 10 days later after the first measure was implemented.  The displayed data indicate that the transition from the PGED regime may be closely connected with the delayed effect of the applied measures. This observation can eventually help to make more accurate predictions for the countries that have not yet reached the PGED regime but have already introduced mitigation measures, i.e. even for countries that do not show any signal of a systematic decay in the data.

\subsection{Countries in the PGED phase close to and past the epidemic peak (05-May-2020)}
Selected countries close to and past the epidemic peak are displayed in Figure~\ref{figure4} and Figures~\ref{figure5}--\ref{figure6}, respectively. Consistent approximate exponential decay in the number of reported active cases in many countries may serve as a sign of a successful strategy against the further spread of the coronavirus. However, due to factors as abatement of the strict mitigation measures, fatigue of following social distancing, or simply due to reintroduction of the virus into the community a further spreading may occur. Such a trend that is demonstrated by a sudden slow down of the exponential decay (Austria, Australia, Vietnam) or even a sign of the next epidemic wave (Azerbaijan, Burkina Faso, Chile, Djibouti, Iran, Iraq, Jordan, Kyrgyzstan, Lebanon, Madagascar) also appears in the reported data, see selected countries in Figure~\ref{figure7}. For Singapore we only model its second epidemic wave.

\subsection{Typical time scale of removal of infected individuals}
The parameter $T_G$ of  PGED characterizes the typical time scale of removal of infected individuals, particularly those who would be eventually tested positive for SARS-CoV-2 infection by a PCR test, i.e., the time period during which an average infected individual can infect others (in susceptible population). A smaller value of $T_G$ correspond to a fast decay after a country reaches its epidemic peak while large values of $T_G$ indicate very flat epidemic peaks and thus a very slow gradual decay of the active cases.
In practice, $T_G$ is influenced by multiple factors, among them ability to identify, test and quarantine positive cases in the population and  contact-tracing procedures play a prominent role. While a larger and better selected infection testing sample can significantly decrease $T_G$ in countries with a higher degree of epidemic, in countries with a small number of active cases finding and testing a small number of infected in the whole population can be very difficult. Therefore contact-tracing and a prevention of an import of  infections from other countries can be the key measures to lower the value of $T_G$.
See Fig.~\ref{figure11} for the graphical display of the sorted values of $T_G$ for countries that are at or beyond the epidemic peak. Countries classified as undergoing the second wave are not included in the plot as a presence of apparent second wave may be eventually a sign of spurious data. Note that the countries that are currently close to their epidemic peak have higher values of $T_G$ than countries that are already further in the decay phase (with the exception of Jamaica).
The data suggest that countries with a small values of $T_G$ (Mauritius$^\ast$, Iceland, Ireland, Australia, Austria, New Zealand) are very efficient in testing and isolation of the individuals who will be tested positive thus preventing them for further spreading of the infection. On the other hand, the data for countries with large values of $T_G$ (Slovenia, Norway, Uruguay) do not show an indication of an efficient testing and isolation of infected (or they may not properly report recovered cases data). However, this interpretation needs to be taken with a caution with regard to the note of precaution in Section~\ref{s:NP}. Particularly, the interpretation of $T_G$ in countries marked with $^\ast$ in Fig.~\ref{figure11} and Tab.~\ref{tab1}--\ref{tab3} can be influenced by the special adjustment of the time series mentioned in previous sections.

\subsection{PGED model as a predictive tool}
The simple PGED model, i.e., the universal scaling \eq{eq:Vazquez2} and nonlinear fitting of the parameters from the data, 
can be used for as a predictive tool for the number of the reported active cases, particularly in countries in the growth phase. Once again keep in mind the note of precaution we formulated in Section~\ref{s:NP}. No verifiable connection of the number of active cases to the total number of infected in the population was established so far. Therefore all the predictions only concern the reported data.

We present a performance of the predictive capabilities of the PGED model using the available data for eight selected countries (Belgium, Belarus, Czechia, Israel, Italy, Portugal, Switzerland, and US) that are in different epidemic phases and also display a variable accuracy of predictions.  
Testing was performed by comparing the values predicted by the PGED model (based on the incomplete data in which we have removed up to 15 data points  from the end of the time series) to the withheld data. For each choice of the number of withheld days we have calculated the 95\% confidence interval for the inferred parameters by MATLAB\textcopyright \ 
function {\tt nlpredci.m}.  Particularly, we were interested in the confidence intervals for the location of the epidemic peak and the number of active cases at the peak. Predictive power of the model can be visualized by plotting the bounding boxes corresponding to the confidence interval around the inferred location of the epidemic peak. 
A good model should provide a consistent position of the confidence intervals with smaller boxes indicating a large degree of certainty in the predictions. 
Note that the analysis using the bounding boxes can be considered also a study of a sensitivity of the fit to the data.

The countries in Fig.~\ref{figure8} are among those past the inflection point,  at the epidemic peak or past the peak. In the latter case we withheld sufficiently many data points so that the estimates would be nontrivial.  Overall, we have found that the short-term prediction (up to 1-2 weeks prior to the peak) tends to predict the location and value of the peak relatively well (Belgium, Portugal, Israel, Switzerland), however, the confidence intervals may be quite large in case not enough data beyond the inflection point is available (Belarus). When the inflection point (and the PGED regime) has not been reached yet, the information about an eventual exponential decay is not directly detectable in the data and the location and the value at the peak thus cannot be estimated, see the remark at the end of Section~\ref{s:3P} for a discussion of a possible prediction improvement for countries in the PG phase, i.e., before they reach the PGED regime. We also illustrate two common situations: while
for US, Italy and Portugal the prediction with less information underestimates the severity of the infection,  for Czechia a forecast overestimates with less data. However, in both cases the fits were changing monotonically with the number of data points included in the analysis. Note that some of these trends may be due to variation in testing procedures and protocols. 

In Fig.~\ref{figure10} we also show how 95\% confidence intervals can be calculated for the whole future data trajectory. This is not straightforward as the 95\% confidence intervals for the estimated parameters are not independent. However, using the covariance structure of the inferred parameters it is possible to sample the parameters from the multivariate normal distribution and display the confidence intervals systematically for all times, as shown in the figure.

\subsection{Application of the PGED model for data predictions}
The presented model can be also visualized using the web based tool \cite{LukasCOVID}. It allows a general public to explore the data for various countries, including validation of the model predictions.
Using the PGED model we have also successfully constructed a prediction on March 30, 2020, for the reported COVID-19 data in Slovakia that estimated a epidemic peak of about 1000 active cases in early May and that very closely matched the observed data, see the reference in the national media \cite{HNCOVID}. At that point in time the prediction differed by orders of magnitude from the predictions of compartment based models. Subsequently, the PGED model was incorporated into the main epidemic (SIR type) model in Slovakia maintained by the analytic unit of the Ministry of Health and that serves as a reference tool for the government crisis management team decision making during the COVID-19 outbreak in Slovakia \cite{IZP}.

\subsection{Analysis of next epidemic wave}
\begin{figure}
\centering
\includegraphics[width=0.8\textwidth]{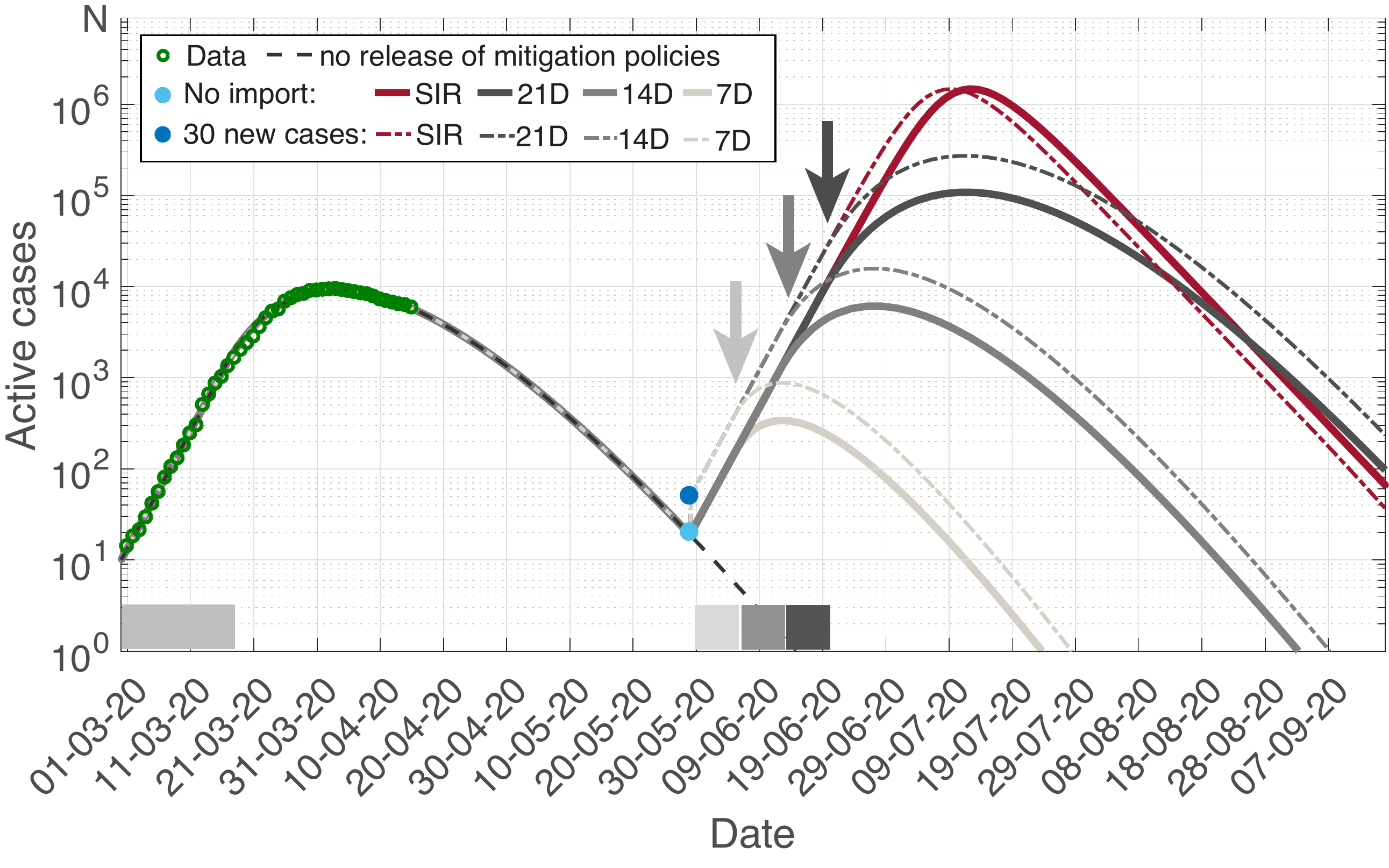}
\caption{Influence of an early intervention on an extent of the next epidemic wave. Data from Austria (\cite{data} from April 15, 2020) are fitted by a combination of  EG (SIR model) and PGED regime (March 1 -- April 15) with a set transition on March 16.  Scenario I (solid lines): 90 days after the epidemic onset mitigation policies are completely lifted while there are still 20 infected individuals in the population. At that point the SIR dynamics restarts. The mitigation policies are reintroduced after 7, 14, and 21  days. Secenario II: In addition 30 more  infected individuals are introduced 90 days after the epidemic onset (dashed lines). Shaded regions indicate the times of duration of the EG regime. 
\label{figure9}}
\end{figure}

\begin{figure}
\centering
\includegraphics[width=1\textwidth]{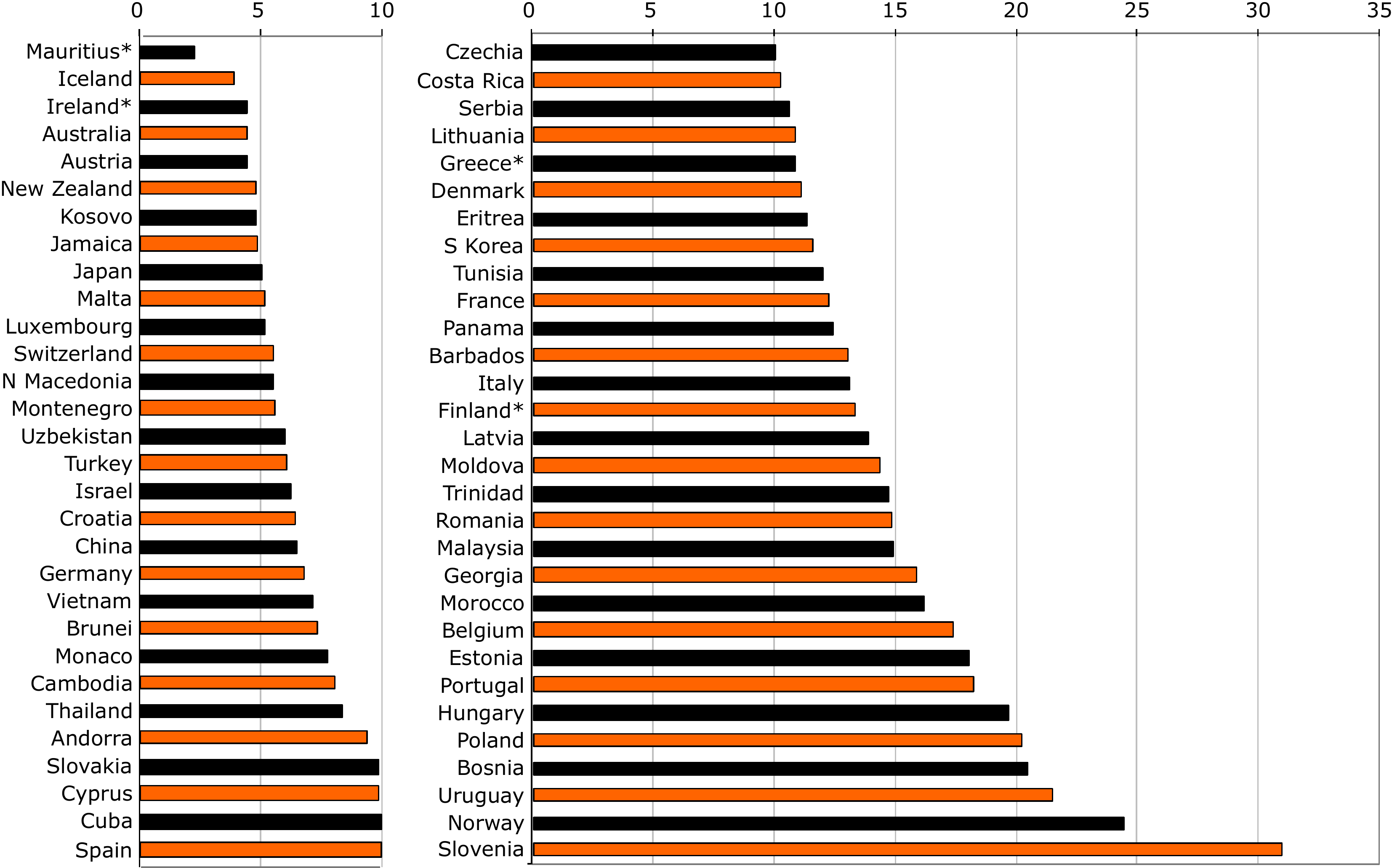}
\caption{The inferred values of the parameter $T_G$ of PGED regime for countries close or beyond their epidemic peak (except those observing an apparent second epidemic wave). Lower $T_G$ corresponds to efficient identification, testing and isolation/removal of infected. Stars indicate modification of the data to account for data reporting irregularities, see the text.
\label{figure11}}
\end{figure}

\begin{figure}
\centering
\includegraphics[width=\textwidth]{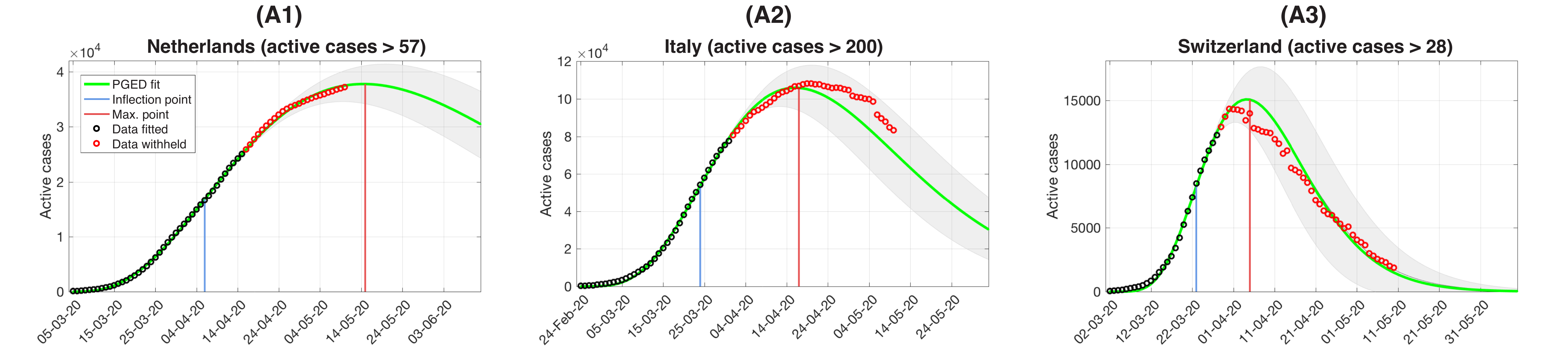}
\caption{Confidence regions for the all future times for selected countries. The data (black) are used for inference of the PGED parameters $P$, $\alpha$ and $T_M$ and their covariance matrix. The best fit is displayed (solid green line).  Symmetric 95\% confidence intervals obtained by sampling parameters from the mutivariate normal distribution with the same covariance structure are displayed as shaded regions. 
\label{figure10}}
\end{figure}

The state of exponential decay of the infected population is often viewed by policy makers as the ultimate goal. However, without reaching the state of  herd immunity the epidemiological situation is unstable with respect to secondary infections caused by rare infected individuals, new imported cases, and related superspreading events. 
We illustrate such a situation in the numerical example in Figure~\ref{figure9}. As an example we consider the reported data in Austria (over the period March 1 -- April 15).
For simplicity we match the data using the EG regime first (using the SIR model with inferred parameters $\beta$ and $\gamma$). The simulation is initialized on March 1 (14 infected, 0 recovered, total population approx.~8.9 mil.). SIR dynamics is applied for the first 16 days reflecting the lack of measures in the early stages of the infection spread (note that the measures reflect in the reported data with a delay). After 16 days we match the rest of the data with PGED model (with inferred parameters $P$, $\alpha$ and $T_M$ and a continuous $R_0$; the values are similar to the parameters for Austria reported in Table~\ref{tab1}). We then continue PGED model until the  May 31 (90 days after the considered initial date). The number of recovered during PGED regime period is calculated from the \eq{Req} and the number of susceptible as the complement of infected and recovered in the population. The remaining population of infected individuals is estimated to be 20 on May 31. 

In the studied scenario we lift the mitigation measures completely on May 31. The dynamics then returns back to the standard SIR model and undergoes an EG phase. 
We study the impact of an early detection of the emerging situation (upcoming second wave) and consequent implementation of mitigation measures. We considered three alternatives: mitigation measures fully implemented after 1, 2, and 3  weeks (see shaded regions in Fig.~\ref{figure9}). We observe that an early implementation of the mitigation measures dramatically reduces the next epidemic peak. A qualitatively similar progress can be seen in case of imported infections (we add 30 new infected cases at the time of released mitigation measures). The numerical results indicate how essential is to implement mitigation measures as early as possible, which requires efficient tools for an early detection of infected individuals. This example shows how the very simple PGED model can be used for analytics of the COVID-19 pandemic in individual countries.

\section{Discussion and Perspectives}
Reported data on COVID-19 display systematically identifiable regimes -- exponential growth, polynomial growth, and polynomial growth with exponential decay.  The observed universal scaling is a bit surprising as the pandemic mitigation and social distancing measures, the testing procedures and protocols, and many other aspects, vary significantly from one country to another. Nevertheless, the scaling appears to be a strong attractor of the reported active cases dynamics globally. An important feature of PG and  PGED regime is that they both contribute to a slowdown of the epidemic growth in the reported data compared to expected dynamics driven by the SIR model.  Note that we have only considered the active cases data but our preliminary data analysis confirms that the PG trends are present in the reported deaths and where available also in reported hospitalizations. 
Therefore we conjecture that the observed transition between the different regimes is comparable to phase transitions in physics, thus one expects that the universal scalings in data are a consequence of some unidentified fundamental properties related to the pandemic. 
Lack of the reliable and detailed data that would allow to discriminate between their eventual sources and a high complexity of the studied  system that involves the virus/disease (its medical, chemical, and physical properties), behavior of individuals in population, and enforcement of the mitigation measures (see \cite{BSImmun} for a summary of some related questions) 
do not allow to identify underlying factors for the observed PG and PGED regimes. Here we only list eventual candidates (or their combinations):
\begin{itemize}
\item
Significant changes in the effective contact network  (social distancing and other mitigation measures)  including low infection transmission probability in majority of contacts due to imposed safety measures (personal protection items as face masks, gloves, disinfectants, etc.).
\item
Limitations of testing procedures and selection of the sample used for testing, including high level of uncertainty in test sensitivity (related to limit of detection and difficulties with sample collection) and specificity of all types of tests, over- and undersampling of various groups in testing, and failure to identify and test asymptomatic carriers; delays in test reporting. 
\item
Limited understanding of details of infection spread mechanisms, particularly the role of individual and temporal variation of viral load in infected individuals and their ability of infect others, related to various clinical stages of the disease; a lack of understanding of mechanisms of superspreading events.
\end{itemize}
Note that the observed PG a PGED regimes are not in agreement with the traditional SIR type models that typically form a base for pandemic spread predictions published in the media unless their parameters are modified from their expected values, particularly a total population is decreased to a significantly lower effective total population. 
Therefore we conclude that  although this work does not provide understanding of the full extent of the pandemic 
as it only models the reported data, it still may provide a useful source 
for decision making, for a comparison of different countries, or for economical predictions by governments, epidemiologists, and economists. 

\section*{Acknowledgments}
This work has been supported by the Slovak Research and Development Agency under the contract no.~APVV-18-0308 (RK) and by the Scientific Grant Agency of the Slovak Republic under the grants no.~1/0755/19 and 1/0521/20. The authors would like to thank Vlado Bo{\v z}a, Luk{\'a}{\v s} Pol{\'a}{\v c}ek, Michal Burger and the modelling team of Institute of Health Policy for their useful comments and help. Particular thanks goes to Robert Ziff for an inspiration and Charlie Doering for pointing us in the right direction.

\end{document}